\documentclass[twocolumn,superscriptaddress,prl]{revtex4-1}
\usepackage{braket}
\usepackage{amsmath,amssymb}
\usepackage{tikz}
\usepackage{graphicx}
\usepackage{color}
\usepackage[colorlinks,bookmarks=false,citecolor=blue,linkcolor=red,urlcolor=blue]{hyperref}
\usepackage{times}

\begin{document}

\title{Entanglement and thermodynamics after a quantum quench in integrable systems} 

\author{Vincenzo Alba}
\author{Pasquale Calabrese}
\affiliation{International School for Advanced Studies (SISSA),
Via Bonomea 265, 34136, Trieste, Italy, 
INFN, Sezione di Trieste}


\begin{abstract}
Entanglement and entropy are key concepts standing at the foundations of quantum and
statistical mechanics. Recently, the study of quantum quenches
revealed that these concepts are intricately intertwined. Although the unitary
time evolution ensuing from a pure state maintains the system  at zero entropy,
local properties at long times are captured by a statistical ensemble with non zero
thermodynamic entropy, which is the entanglement accumulated  during
the dynamics.
Therefore, understanding the  entanglement evolution unveils how
thermodynamics emerges in isolated systems. Alas, an exact computation
of the entanglement dynamics was available so far only for non-interacting 
systems, while it was deemed unfeasible for interacting ones.
Here we show that the standard quasiparticle picture of the entanglement
evolution, complemented with integrability-based knowledge of the steady state and its
excitations, leads to a complete understanding of the entanglement dynamics 
in the space-time scaling limit. We thoroughly check our result for the paradigmatic 
Heisenberg chain.
\end{abstract}


\maketitle

Since the early days of quantum mechanics,
understanding how statistical ensembles arise from the unitary time evolution 
of an isolated quantum system has been a fascinating question \cite{v-29,eth0,deutsch-1991,
srednicki-1994,rigol-2008,rigol-2012,dalessio-2015}. 
A widely accepted mechanism is that while the entire system remains in a pure state, the reduced density matrix of an 
arbitrary finite compact subsystem attains a long time limit that can be described by a statistical ensemble (see, e.g., Ref. \cite{essler-2016}).
In the last decade ground-breaking experiments with cold atoms~\cite{kinoshita-2006,hofferberth-2007,trotzky-2012,gring-2012,cheneau-2012,
langen-2013,meinert-2013,fukuhara-2013,langen-2015,islam-2015,kaufman-2016}
simulated with incredible precision the unitary time evolution of many-body 
quantum systems, reviving the interest in this topic.
The simplest out-of-equilibrium protocol in which these ideas can be theoretically 
and experimentally tested is the quantum quench~\cite{calabrese-2006,polkovnikov-2011}: 
A system is prepared in an initial state $|\Psi_0\rangle$, typically the ground 
state of a local Hamiltonian ${H}_0$, and it  evolves with a 
many-body Hamiltonian ${H}$. At asymptotically long times, physical 
observables relax to stationary values, which for generic systems are described by 
the Gibbs (thermal) ensemble~\cite{deutsch-1991,srednicki-1994,rigol-2008,rigol-2012,dalessio-2015}, 
whereas for {\it integrable} systems a Generalized Gibbs Ensemble (GGE) has to 
be used~
\cite{rigol-2007,collura-2013a,bastianello-2016,vernier-2016,alba-2015,
cramer-2008,barthel-2008,cramer-2010,calabrese-2011,calabrese-2012,cazalilla-2006,cazalilla-2012a,
sotiriadis-2012,collura-2013,fagotti-2013,kcc14,sotiriadis-2014,essler-2015,essler-2016,vidmar-2016,
calabrese-2016,ilievski-2015a,cardy-2015,gogolin-2015,sotiriadis-2016}.

Although these results suggest a spectacular compression of the amount of  
information needed to describe steady states, state-of-the-art numerical methods, 
such as the time-dependent Density Matrix Renormalization Group~\cite{white-2004,
daley-2004,uli-2005,uli-2011} (tDMRG), can only access the short-time dynamics. 
Physically, the origin of this conundrum is the fast growth of the entanglement 
entropy $S\equiv- \textrm{Tr}_A\rho_A\ln\rho_A$, with $\rho_A$ being the reduced 
density matrix of an interval $A$ of length $\ell$ embedded in an infinite system. 
It is well-understood that $S$ grows linearly with the time after the quench~\cite{calabrese-2005}. 
This implies an exponentially increasing amount of information manipulated 
during typical tDMRG simulations. 
Remarkably, the entanglement dynamics has been successfully observed in a very recent 
cold-atom experiment~\cite{kaufman-2016}. 

\begin{figure}[t]
\includegraphics*[width=0.82\linewidth]{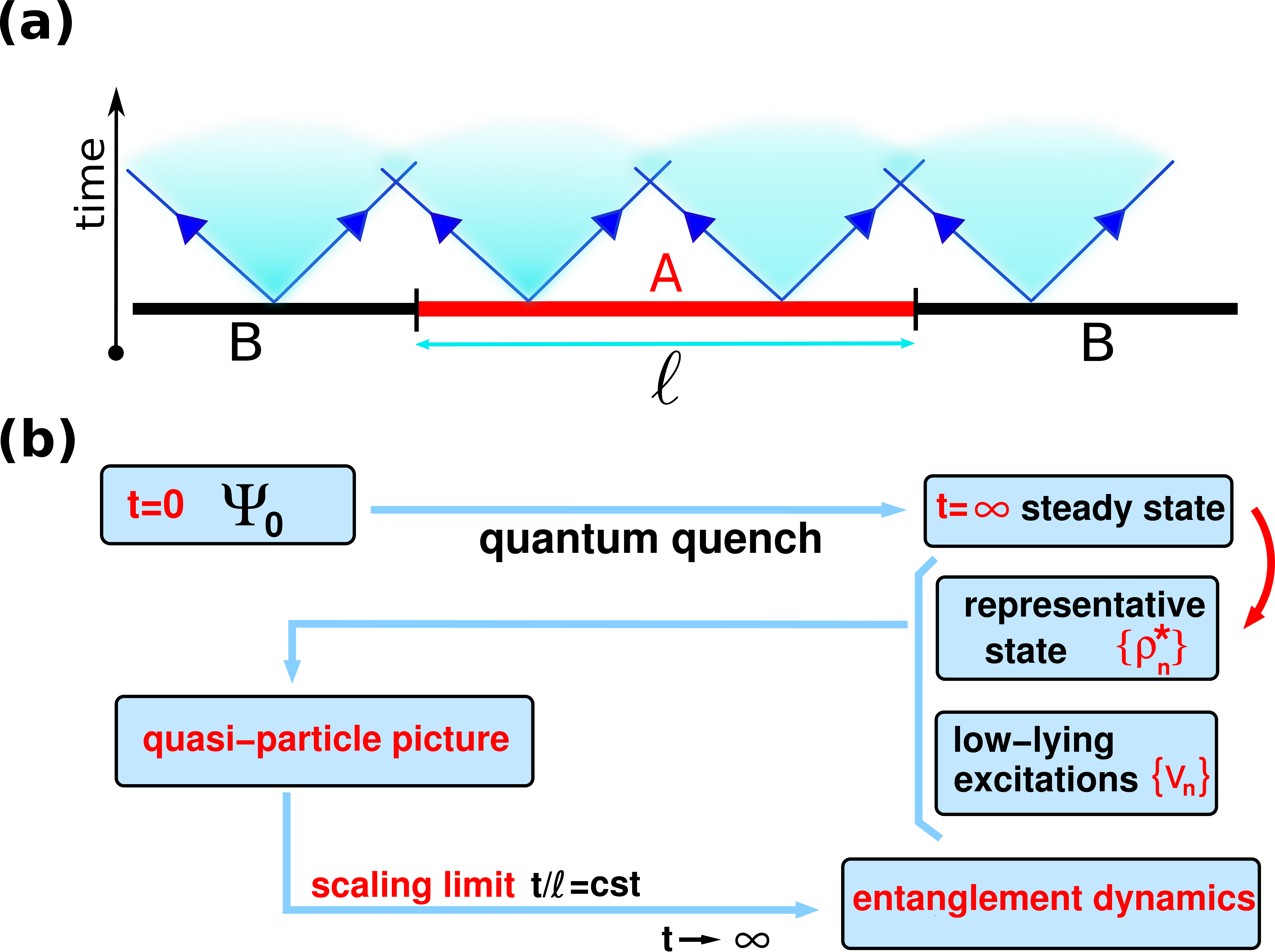}
\caption{Entanglement dynamics after a quantum quench: Theoretical scheme. 
 (a) Quasiparticle picture. Full lines denote quasiparticles 
 with maximum velocity emitted in the initial state. Shaded cones: Halo of 
 slower quasiparticles. (b) Main steps to calculate the entanglement 
 dynamics using Bethe ansatz and the quasiparticle picture. 
}
\label{fig0}
\end{figure}

In this Letter, using a standard quasiparticle picture~\cite{calabrese-2005}, 
we show that in integrable models the steady state and its low-lying excitations encode 
sufficient information to reconstruct the entanglement dynamics up to  short times. 
According to the quasiparticle picture~\cite{calabrese-2005}, the pre-quench initial state 
acts as a source of pairs of quasiparticle excitations. Let us 
first assume that there is only one type of quasiparticles identified by their 
quasimomentum $\lambda$, and moving with velocity $v(\lambda)$. 
While quasiparticles created far apart from each other are incoherent, those emitted 
at the same point in space are entangled. As these propagate ballistically throughout the system, 
larger regions get entangled. 
At time $t$, $S(t)$ is proportional to the total number of quasiparticles that,
emitted in pairs from the same point, reach one the subsystem $A$ and the other its complement (see 
Fig.~\ref{fig0} (a)). Specifically, one obtains 
\begin{equation}
\label{semi-cl}
S(t)\propto 2t\!\!\!\!\int\limits_{\!2|v|t<\ell}\!\!\!\!d\lambda v(\lambda)f(\lambda)+
\ell\!\!\!\!\int\limits_{2|v|t>\ell}\!\!\!\!d\lambda f(\lambda), 
\end{equation}
where $f(\lambda)$ depends on the production rate of  quasiparticles. 
\eqref{semi-cl} holds in the space-time 
scaling limit $t,\ell\to \infty$ at $t/\ell$ fixed. When a maximum 
quasiparticle velocity $v_M$ exists (e.g., due to the Lieb-Robinson 
bound~\cite{lieb-1972}), for $t\le \ell/(2v_M)$, 
$S$ grows linearly in time because the second term in~\eqref{semi-cl} vanishes. 
In contrast, for $t\gg\ell/(2v_M)$ the entanglement 
is extensive, i.e., $S\propto\ell$. 
This light-cone spreading of entanglement has been confirmed analytically 
only in free models~\cite{fagotti-2008,ep-08,nr-14,coser-2014,cotler-2016,buyskikh-2016}, numerically in several studies 
(see e.g.~\cite{de-chiara-2006,lauchli-2008,kim-2013}), in the holographic 
framework~\cite{hrt-07,aal-10,aj-10,allais-2012,callan-2012,ls-14,bala-2011,liu-2014}, and in a recent 
experiment~\cite{kaufman-2016}. 

%
%

\paragraph*{\bf Results.} 
In a generic interacting integrable model, there are different species of stable
quasiparticles, corresponding to bound states of an arbitrary number of elementary 
excitations. Integrability implies that different types of quasiparticles must 
be treated independently. It is then natural to conjecture that 
\begin{equation}
\label{conj}
S(t)= \sum_n\Big[ 2t\!\!\!\!\!\!\int\limits_{\!2|v_n|t<\ell}\!\!\!\!\!\!
d\lambda v_n(\lambda)s_n(\lambda)+\ell\!\!\!\!\!\!\int\limits_{2|v_n|t>\ell}\!\!\!\!\!\!
d\lambda s_n(\lambda)\Big],
\end{equation}
where the sum is over the types of particles $n$, $v_n(\lambda)$ is their velocity, 
and $s_n(\lambda)$ their entropy. To give predictive power to~\eqref{conj}, in the 
following we show how to determine $v_n(\lambda)$ and $s_n(\lambda)$ in the Bethe ansatz 
framework for integrable models. 

The eigenstates of Bethe ansatz solvable models are in correspondence with a 
set of pseudomomenta (rapidities) $\lambda$. In the thermodynamic limit, these 
form a continuum. One then introduces the particle densities $\rho_{n,p}(\lambda)$, 
the hole (i.e., unoccupied rapidities) densities $\rho_{n,h}(\lambda)$, and the total 
densities $\rho_{n,t}(\lambda)=\rho_{n,p}(\lambda)+\rho_{n,h}(\lambda)$. Every set 
of densities identifies a thermodynamic ``macro-state''. This corresponds to an 
exponentially large number of microscopic eigenstates, any of which can be used as a  
``representative'' for the macro-state. 
The total number of representative microstates is $e^{S_{YY}}$, with $S_{YY}$ 
the thermodynamic Yang-Yang  entropy  of the macrostate 
\begin{multline}
\label{s-yy}
S_{YY}= s_{YY}L=L \sum_{n=1}^{\infty}\int\!d\lambda
[\rho_{n,t}(\lambda) \ln \rho_{n,t}(\lambda) \\ - \rho_{n,p}(\lambda) \ln
\rho_{n,p}(\lambda)- \rho_{n,h} (\lambda)\ln \rho_{n,h}(\lambda)]
\\ \equiv L \sum_{n=1}^{\infty}\int\!d\lambda s_{YY}^{(n)}[\rho_{n,p},
\rho_{n,h}](\lambda).
\end{multline}
Physically, $S_{YY}$ corresponds to the total number of ways of assigning the 
quasimomentum label to the particles, similar to free-fermion models. 
%
\begin{figure}[t]
\includegraphics*[width=0.98\linewidth]{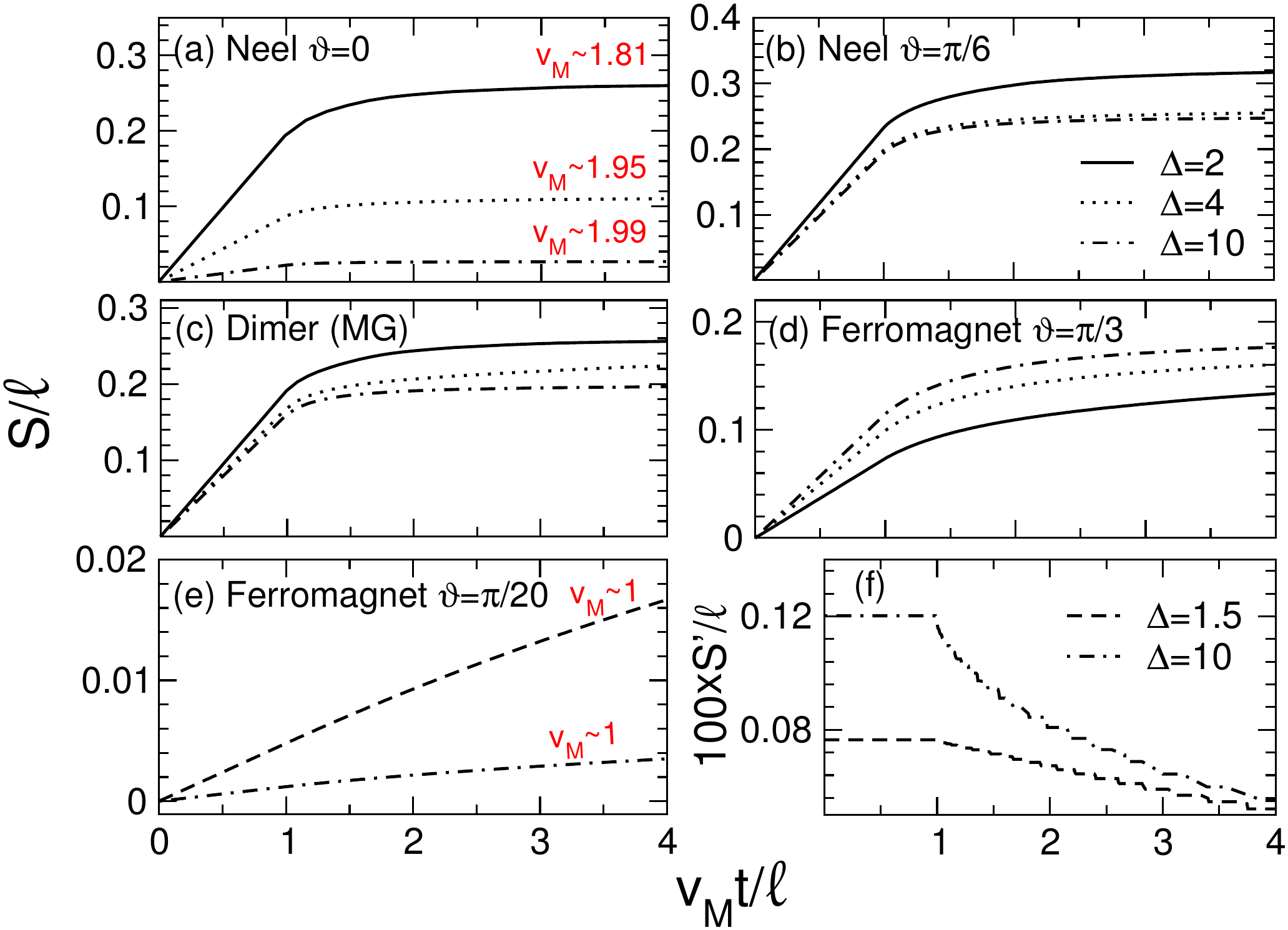}
\caption{Analytical predictions for the $XXZ$ chain. 
 Entanglement  entropy per site $S/\ell$ versus $v_M t/\ell$, with 
 $v_M$ the maximum velocity. Different panels correspond to the 
 different initial states and different lines to different $\Delta$. 
 For $\Delta\to\infty$, $S\to 0$ for the N\'eel quench, whereas it 
 saturates in the other cases. Note in (e) the substantial  
 entanglement increase for $v_Mt/\ell>1$. Panel (f): The numerical 
 derivative $S'(v_Mt/\ell)\times100$ for the quench in (e). 
}
\label{fig2}
\end{figure}

\begin{figure}[b]
\includegraphics*[width=0.98\linewidth]{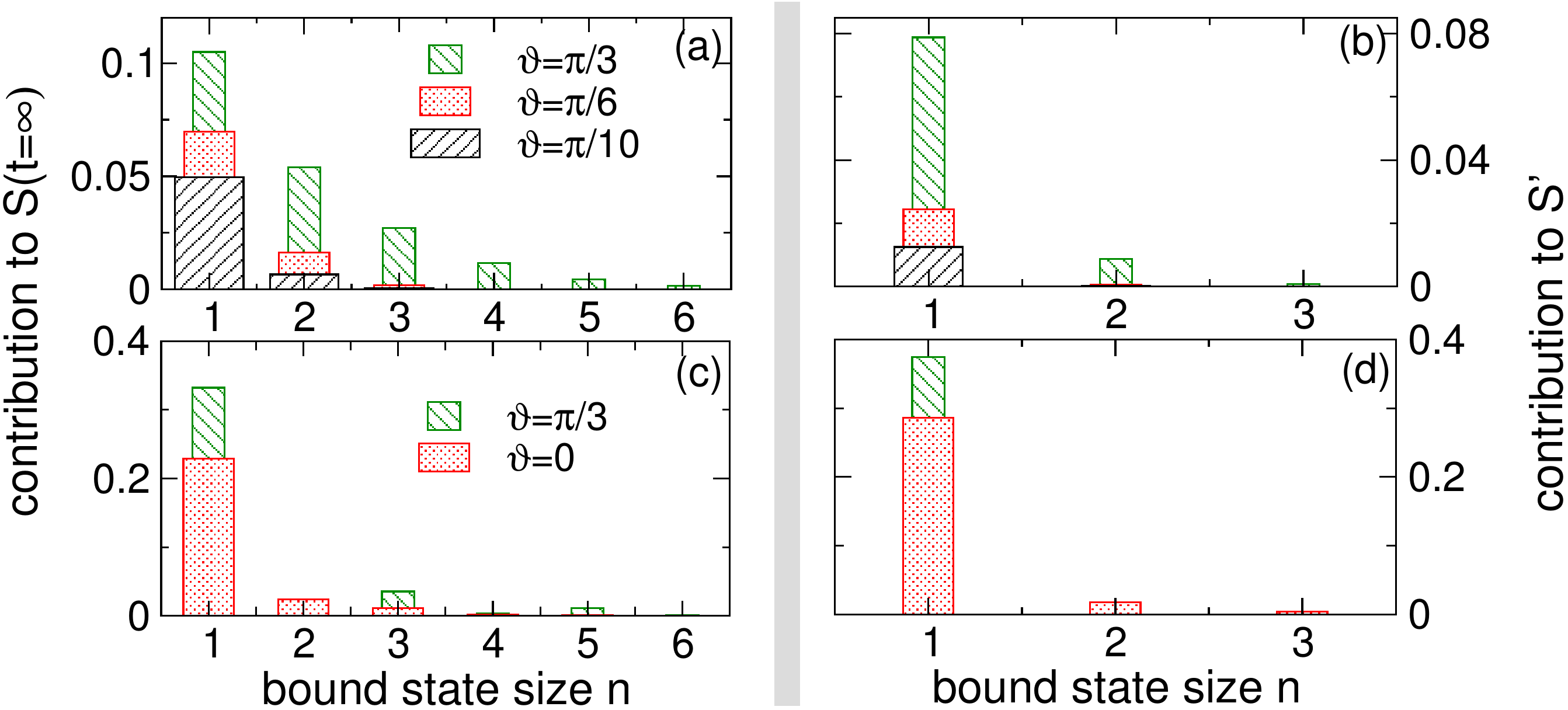}
\caption{Bound-state contributions to the entanglement dynamics. On the $x$-axis 
 $n$ is the bound-state size. (a)(b)  Quench from the tilted ferromagnet. Panel (a): 
 Bound-state contributions to steady-state entropy density (second term in ~\eqref{conj}). 
 Panel (b): Bound-state contributions to the slope of the entanglement growth 
 for $t<\ell/v_M$ (first term in~\eqref{conj}). Different histograms denote 
 different tilting angles $\vartheta$. All the data are for $\Delta=2$. (c)(d): 
 same as in (a)(b) for the quench from the tilted N\'eel state. 
}
\label{fig3}
\end{figure}
%

%
\begin{figure*}[t]
\includegraphics*[width=0.93\linewidth]{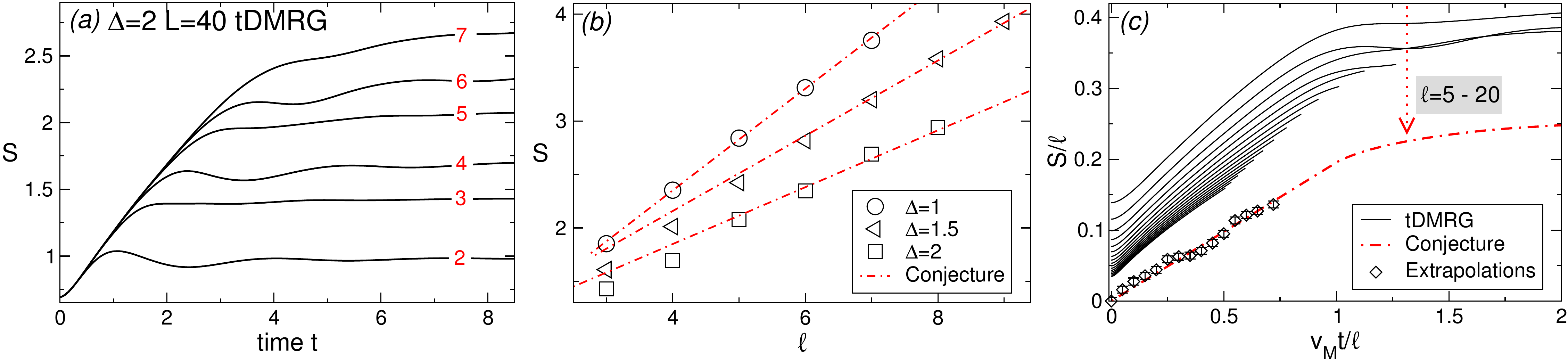}
\caption{Comparison with numerical simulations. Entanglement entropy 
 dynamics after the quench from the N\'eel  state in the $XXZ$ chain.
 (a) tDMRG results for a chain with $L=40$ sites and $\Delta=2$. 
 Different curves correspond to different subsystem sizes $\ell$ (accompanying numbers). 
 (b) The entropy saturation values (tDMRG results at $t\approx 8$)  as a function of the 
 block length $\ell$, for several $\Delta$. 
 The dashed-dotted lines is the  conjectured volume-law behavior $S\propto s^*_{YY}\ell$ (cf.~\eqref{s-yy}). 
 (c) The scaling limit: $S/\ell$ plotted versus $v_Mt/\ell$. The continuous curves are the tDMRG 
 results for $\ell=5- 20$.  The diamonds are numerical extrapolations to the thermodynamic limit. 
 The dashed-dotted line is the conjecture~\eqref{conj}. 
}
\label{fig4}
\end{figure*}

In the Bethe ansatz treatment of quantum quenches~\cite{caux-2013,caux-2016,iqdb-15}, 
local properties of the post-quench stationary state are described by 
a set of densities $\rho_{n,p}^*(\lambda)$ and $\rho_{n,h}^*(\lambda)$.
Calculating these densities is a challenging task that has been performed only in a few 
cases~\cite{wouters-2014A,pozsgay-2014A,bse-14,dwbc-14,bucciantini-15,ac-16,pce-16,piroli-2016,mestyan-2015,brockmann-2014,piroli-2016-a,pvcr-16,bpc-16}. 
From the densities, 
the thermodynamic entropy of the stationary ensemble~\eqref{s-yy}  is $s_{YY}[\rho_{n,p}^*,\rho_{n,h}^*]
(\lambda)$. Physically, this reflects a generalized microcanonical ensemble for quenches, in which 
all the microstates corresponding to the macrostate have the same probability. 

We now present our predictions for the entanglement dynamics (Fig.~\ref{fig0} (b) 
gives a survey of our theoretical scheme). 
First, in the stationary state the density of thermodynamic entropy coincides with 
that of the entanglement entropy in~\eqref{conj}, as it has been shown analytically for 
free models~\cite{fagotti-2008,ckc-14,kbc-14}. 
This implies that $s_n(\lambda)=s_{YY}[\rho_{n,p}^*,\rho_{n,h}^*](\lambda)$. Moreover, it is 
natural to identify the entangling quasiparticles 
in~\eqref{conj} with the low-lying  excitations around the stationary 
state $\rho^*$. Their group velocities $v_n$ depend on the stationary state, because 
the interactions induce a state-dependent dressing of the excitations. 
These velocities $v_n$ can be calculated by Bethe ansatz techniques~\cite{bonnes-2014} (see Supplementary material). 

To substantiate our idea we focus on the spin-$1/2$ anisotropic Heisenberg ($XXZ$) chain,   
considering quenches from several low-entangled initial states, namely the tilted N\'eel 
state, the Majumdar-Ghosh (dimer) state, and the tilted ferromagnetic state (see the 
Methods paragraph). 
For these initial states the densities $\rho^*_{n(h),p}$ are known analytically 
\cite{wouters-2014A,pozsgay-2014A,piroli-2016-a}.


Fig.~\ref{fig2} summarizes the expected entanglement dynamics 
in the space-time scaling limit, plotting $S/\ell$ versus 
$v_Mt/\ell$. Interestingly, $S/\ell$ is always smaller than $\ln2$, i.e., 
the entropy of the maximally entangled state. 
For the N\'eel quench, since the N\'eel state becomes the ground state 
of~\eqref{ham-xxz} in the limit $\Delta\to\infty$, $S/\ell\approx\ln(\Delta)/\Delta^2$ 
vanishes, whereas it saturates for all the other quenches. 
For the Majumdar-Ghosh state one obtains $S/\ell=-1/2+\ln2$ at $\Delta\to\infty$. 
For the tilted ferromagnet with $\vartheta\to0$ 
(Fig.~\ref{fig2} (e)), $S/\ell$ is small at any $\Delta$, reflecting that the ferromagnet 
is an eigenstate of the $XXZ$ chain. Surprisingly, the linear growth seems to extend 
for $v_Mt/\ell>1$. However, $dS/dt$ (Fig.~\ref{fig2} (f)) is flat only for 
$v_{M}t/\ell\le 1$, which signals true linear regime only for $v_Mt/\ell\le1$. This peculiar  
behavior is due to the large entanglement contribution of the slow quasiparticles. 
In Fig.~\ref{fig3} we report the bound-state resolved contributions to the entanglement 
dynamics. Panel (a) and (c) focus on the steady state entropy (second term 
in~\eqref{conj}), while panels (b) and (d) show the bound-state contributions to the 
slope of the linear growth (first term in~\eqref{conj}). The contribution of  
the bound states, although never dominant, is crucial to ensure accurate predictions.

\begin{figure}[t]
\includegraphics*[width=0.93\linewidth]{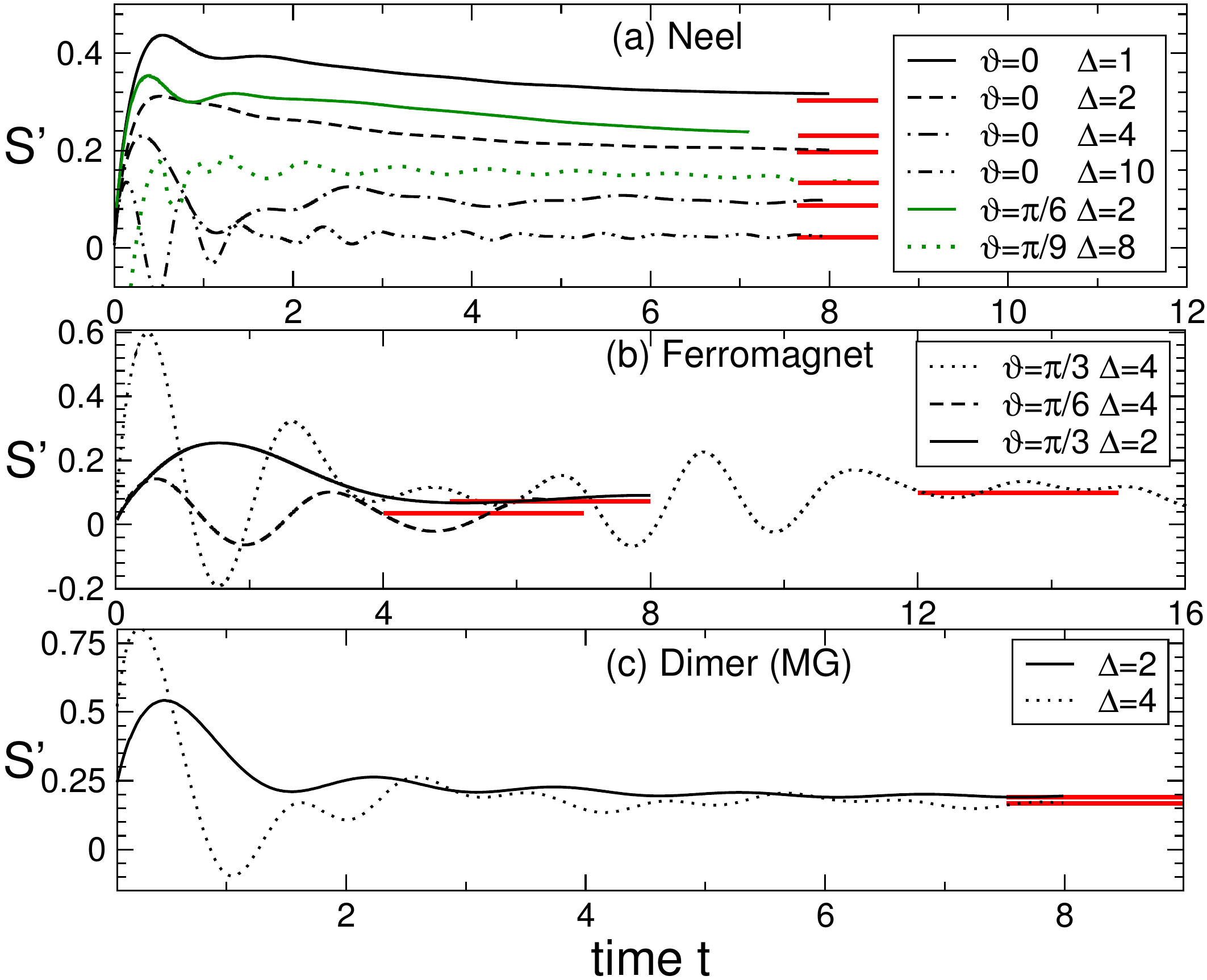}
\caption{Comparison with numerical simulations: The short-time regime.
 Derivative $S'(v_Mt)$ as a function of time. 
 Different panels correspond to different initial states.  $\vartheta$ is the 
 tilting angle. In each panel the different curves are iTEBD results for 
 different $\Delta$. The horizontal lines are the conjecture~\eqref{conj}.  
}
\label{fig5}
\end{figure}

Fig.~\ref{fig4} (a) shows tDMRG results for $S(t)$ for the quench from the symmetrized N\'eel state 
$(\left|\uparrow\downarrow\uparrow\dots\right\rangle+ \left|\downarrow\uparrow\downarrow\dots\right\rangle)/\sqrt2$. 
The data are for the {\it open} $XXZ$ chain, and subsystems starting from the chain boundary. The 
qualitative agreement with~\eqref{conj} is apparent. 
Fig.~\ref{fig4} (b) reports the steady-state entanglement entropy 
as a function $\ell$ (data at $t\approx 8$ in (a)). The volume-law $S\propto\ell$ 
is visible. The dashed-dotted lines are fits to $S\propto s^*_{YY}\ell$, supporting the 
equivalence between entanglement and thermodynamic entropy. Fig.~\ref{fig4} (c) 
focuses on the full time dependence, plotting $S/\ell$ versus $v_Mt/\ell$. 
The dashed-dotted line is~\eqref{conj} with $t\to t/2$, due to the open boundary 
conditions~\cite{de-chiara-2006}. Deviations from~\eqref{conj} due to the finite $\ell$ 
are visible. The diamonds are numerical extrapolations to the thermodynamic limit. 
The agreement with~\eqref{conj} is perfect. 
Finally, we provide a more stringent check of~\eqref{conj}, focusing on the linear 
entanglement growth. Fig.~\ref{fig5} shows infinite time-evolving block decimation
(iTEBD) results in the thermodynamic limit for $S'(v_Mt)$, with $S'(x)\equiv dS(x)/dx$
taken from Ref. \cite{fagotti-2014}. 
For all the quenches, the agreement with~\eqref{conj} (horizontal lines) is spectacular. 

%
%

\paragraph*{\bf Conclusions.}
The main result of this Letter is the analytical prediction in~\eqref{conj} for the 
time-dependent entanglement entropy after a generic quantum quench in an integrable model. 
We tested our conjecture for several quantum quenches in the $XXZ$ spin chain, although we 
expect~\eqref{conj} to be more general. Further checks of~\eqref{conj} (e.g., for the 
Lieb-Liniger gas) are desirable. It would be also interesting to generalize~\eqref{conj} 
to quenches from inhomogeneous initial states,  exploiting the recent analytical results \cite{cdy-16,bertini-2016,dsvc-16}. 
Although we are not able yet to provide an {\it ab initio} derivation of~\eqref{conj}, 
we find remarkable that it is possible to characterize analytically the 
dynamics of the entanglement entropy, while its equilibrium behavior is still 
an open challenge. 
Finally, we believe that~\eqref{conj} represents a deep conceptual breakthrough since  
it shows in a single compact formula the relation between entanglement and thermodynamic 
entropy for integrable models. An analogous description for non integrable systems, where 
quasiparticles  have finite lifetime or dot not exist at all, could 
lead to a deeper understanding of thermalization~\cite{kaufman-2016}.

\paragraph*{\bf Methods.} The anisotropic spin-$1/2$ Heisenberg chain is defined by the Hamiltonian 
\begin{equation}
\label{ham-xxz}
{H}=\sum_{i=1}^L\Big[\frac{1}{2}(S_i^+S^-_{i+1}+S_i^+S_{i+1}^-)+
\Delta\Big(S_i^zS_{i+1}^z-\frac{1}{4}\Big)\Big], 
\end{equation}
where $S_i^\alpha$ are spin-$1/2$ operators, and $\Delta$ is the anisotropy 
parameter. Here we considered as pre-quench initial states the tilted N\'eel 
state $|\vartheta,\nearrow\swarrow\cdots\rangle\equiv e^{i\vartheta\sum_jS_j^y}
\left|\uparrow\downarrow\cdots\right\rangle$, the Majumdar-Ghosh (dimer) state 
$|MG\rangle\equiv((\left|\uparrow\downarrow\right\rangle-\left|\downarrow\uparrow
\right\rangle)/2)^{\otimes L/2}$, and the tilted ferromagnetic state $\left|
\vartheta,\nearrow\nearrow\right\rangle\equiv e^{i\vartheta\sum_j S_j^y}
\left|\uparrow\uparrow\cdots\right\rangle$. The Heisenberg spin chain is the 
prototype of all integrable models. Moreover, for 
all the initial states considered in this work the post-quench steady state can 
be characterized analytically, via the ``macro-state'' densities $\rho^*_{p(h)}$. 
Specifically, 
a set of recursive relations for these densities can be obtained (see Supplementary Material). The group velocities 
of the low-lying excitations around the steady state, i.e. the entangling 
quasiparticles,  are obtained by solving numerically an infinite set of second type 
Fredholm integral equations (details are in the Supplementary Material). 

The numerical data for the post-quench dynamics of the entanglement entropy presented 
in Fig.~\ref{fig4} were obtained using the standard tDMRG~\cite{white-2004,daley-2004,uli-2005,uli-2011} 
in the framework of Matrix Product States (MPS). 
For the implementation we used the ITENSOR library (\url{http://itensor.org/}).
The data presented in Figure~\ref{fig5} are obtained using the iTEBD 
method~\cite{vidal-2007} and are a courtesy of Mario Collura.

\paragraph*{\bf Acknowledgments.}

The authors acknowledge support from the ERC under the Starting Grant 279391 EDEQS.  
This project has received funding fromt the European Union's Horizon 2020 research and 
innovation programme under the Marie Sklodowoska-Curie grant agreement No 702612 OEMBS. 
The iTEBD data presented in Fig.~\ref{fig5} are a courtesy of Mario Collura. We also thank 
Lorenzo Piroli and Eric Vernier for sharing their results before publication.


\section*{\large Supplemental material}

In this Supplementary material we provide some details on:
\\(1) The Bethe ansatz solution of the $XXZ$ chain.
\\(2) The Bethe ansatz treatment of the post-quench steady state.
\\(3) The calculation of the group velocities of the low-lying excitations 
around the steady state. 
\\(4) The entanglement and the mutual information of two disjoint intervals.

\section{Bethe ansatz solution of the $XXZ$ chain} 
\label{ba-sol}

In the Bethe ansatz~\cite{taka-book} solution of the $XXZ$ chain, the eigenstates 
of~\eqref{ham-xxz} in the sector with $M$ down spins (particles) are in correspondence 
with a set of rapidities $\lambda_j$. These are obtained by solving the Bethe equations~\cite{taka-book} 
\begin{equation}
\label{be}
\left[\frac{\sin(\lambda_j+i\frac{\eta}{2})}{\sin(\lambda_j-i\frac{\eta}{2})}\right]^L=
-\prod\limits_{k=1}^M\frac{\sin(\lambda_j-\lambda_k+i\eta)}{\sin(\lambda_j-\lambda_k-i
\eta)},
\end{equation}
where $\eta\equiv\textrm{arccosh}(\Delta)$. 

In the thermodynamic limit the solutions of the Bethe equations~\eqref{be} form 
string patterns in the complex plane. The rapidities forming a $n$-string are 
parametrized as 
\begin{equation}
\lambda^j_{n,\gamma}=\lambda_{n,\gamma}+i\frac{\eta}{2}(n+1-2j)+\delta^j_{n,\gamma}, 
\end{equation}
where $j=1,\dots,n$ labels the different string 
components, $\lambda_{n,\gamma}$ is the ``string center'', and $\delta_{n,\gamma}^j$ 
are the so-called string deviations. Typically, i.e., for the majority of the eigenstates 
of~\eqref{ham-xxz}, one has $\delta_{n,\gamma}^j={\mathcal O}(e^{-L})$, implying that 
the string deviations can be neglected~\cite{taka-book} (string hypothesis). Physically, $n$-strings 
correspond to bound states of $n$ down spins. The string centers 
$\lambda_{n,\gamma}$ are solutions of the Bethe-Gaudin-Takahashi (BGT) 
equations~\cite{taka-book}
\begin{equation}
\label{bgt-eq}
L\theta_n(\lambda_{n,\alpha})=2\pi I_{n,\alpha}+\sum\limits_{(n,\alpha)
\ne(m,\beta)}\Theta_{n,m}(\lambda_{n,\alpha}-
\lambda_{m,\beta}). 
\end{equation}
For $\Delta>1$, one has $\lambda_{n,\gamma}\in[-\pi/2,\pi/2)$. Here we define $\theta_n(
\lambda)\equiv2\arctan[\tan(\lambda)/\tanh(n\eta/2)]$. The scattering phases $\Theta_{n,m}
(\lambda)$ are defined as 
\begin{multline}
\label{Theta}
\Theta_{n,m}(\lambda)\equiv(1-\delta_{n,m})\theta_{|n-m|}(\lambda)+2\theta_{
|n-m|+2}(\lambda)\\+\cdots+\theta_{n+m-2}(\lambda)+\theta_{n+m}(\lambda). 
\end{multline}

Each different choice of the so-called BGT quantum numbers $I_{n,\alpha}\in\frac{1}{2}
\mathbb{Z}$ identifies a different set of solutions of~\eqref{bgt-eq}, and, in turn, a 
different eigenstate of~\eqref{ham-xxz}. The corresponding eigenstate energy $E$ and 
total momentum $P$ are obtained by summing over all the BGT rapidities~\cite{taka-book} 
as $E=\sum_{n,\alpha}\epsilon_n(\lambda_{n,\alpha})$, and $P=\sum_{n,\alpha}z_n(
\lambda_{n,\alpha})$ with 
\begin{equation}
\label{eps}
\epsilon_n(\lambda)\equiv-\frac{\sinh(\eta)\sinh(n\eta)}{\cosh(n\eta)-\cos(2\lambda)}, 
\quad z_n(\lambda_{n,\alpha})=\frac{2\pi I_{n,\alpha}}{L}. 
\end{equation}
Note that $P$ depends only on the $I_{n,\alpha}$. 

\section{The steady state: Bethe ansatz treatment}
\label{steady-state}

Here we provide some details on how to derive the steady-state densities 
$\rho^*_{n,p},\rho^*_{n,h}$ in Bethe ansatz. First, the root densities 
$\rho_{n,p}$  are defined as~\cite{yang-1969}
\begin{equation}
\rho_{n,p}(\lambda)\equiv\lim_{L\to\infty}  \frac1{L(\lambda_{n,\alpha+1}-\lambda_{n,\alpha})}. 
\end{equation}
Instead of working with 
the hole densities $\rho_{n,h}$, it is convenient to define $\eta_n=\rho_{n,h}/
\rho_{n,p}$. For all the initial states considered in this work the corresponding 
steady-state densities $\rho^*_{n,h},\eta^*_{n}$ obey the recursive relations~\cite{brockmann-2014,iqdb-15} 
\begin{align}
\label{eta-rec}
&\eta^*_n(\lambda)=\frac{\eta^*_{n-1}(\lambda+i\frac{\eta}{2})
\eta^*_{n-1}(\lambda-i\frac{\eta}{2})}{1+\eta^*_{n-2}(\lambda)}-1,\\
\label{rho-rec}
& \rho^*_{n,h}(\lambda)=\rho^*_{n,t}(\lambda+i\frac{\eta}{2})+\rho^*_{n,t}
(\lambda)-\rho^*_{n-1,h}(\lambda), 
\end{align}
with $\eta^*_{0}=0$ and $\rho^*_{0,h}=0$. In~\eqref{eta-rec} and \eqref{rho-rec} 
the initial conditions $\rho^*_{1,h}$ and $\eta^*_{1}$ encode the information 
about the pre-quench initial state. For all the quenches considered in this 
work, this initial conditions are known analytically. For instance, for the 
quench from the N\'eel state one has~\cite{brockmann-2014}
\begin{align}
\label{eta-neel}
& \eta_1^*=\frac{2[2\cosh(\eta)+2\cosh(3\eta)-3\cos(2\lambda)
\sin^2(\lambda)]}{[\cosh(\eta)-\cos(2\lambda)][\cosh(4\eta)-\cos(4\lambda)]}\\
\nonumber
& \rho^*_{1,h}=\frac{\theta_1'(\lambda)}{2\pi}\Big(1-\frac{4\cosh^2(\eta)}
{[\theta_1'(\lambda)\sin(2\lambda)]^2+4\cosh^2(\eta)}\Big), 
\end{align}
where $\theta'_1(\lambda)\equiv d\theta_1(\lambda)/d\lambda$ (with $\theta_1(\lambda)$ as defined in~\eqref{Theta}). 
For the dimer state $\rho^*_{1,p}$ and $\eta^*_1$ 
have been calculated in~\cite{mestyan-2015} while for the tilted N\'eel  
and the tilted ferromagnet they have been derived recently~\cite{piroli-2016-a,pvcr-16}. 
Figure~\ref{fig1-supp} (a) shows $\rho^*_{n,p}$ for $n=1,2,3$ for the quench from the 
N\'eel state in the $XXZ$ chain with $\Delta=2$.

\section{Group velocities of the entangling quasiparticles}
\label{velocities}

%
\begin{figure}[t]
\includegraphics*[width=0.93\linewidth]{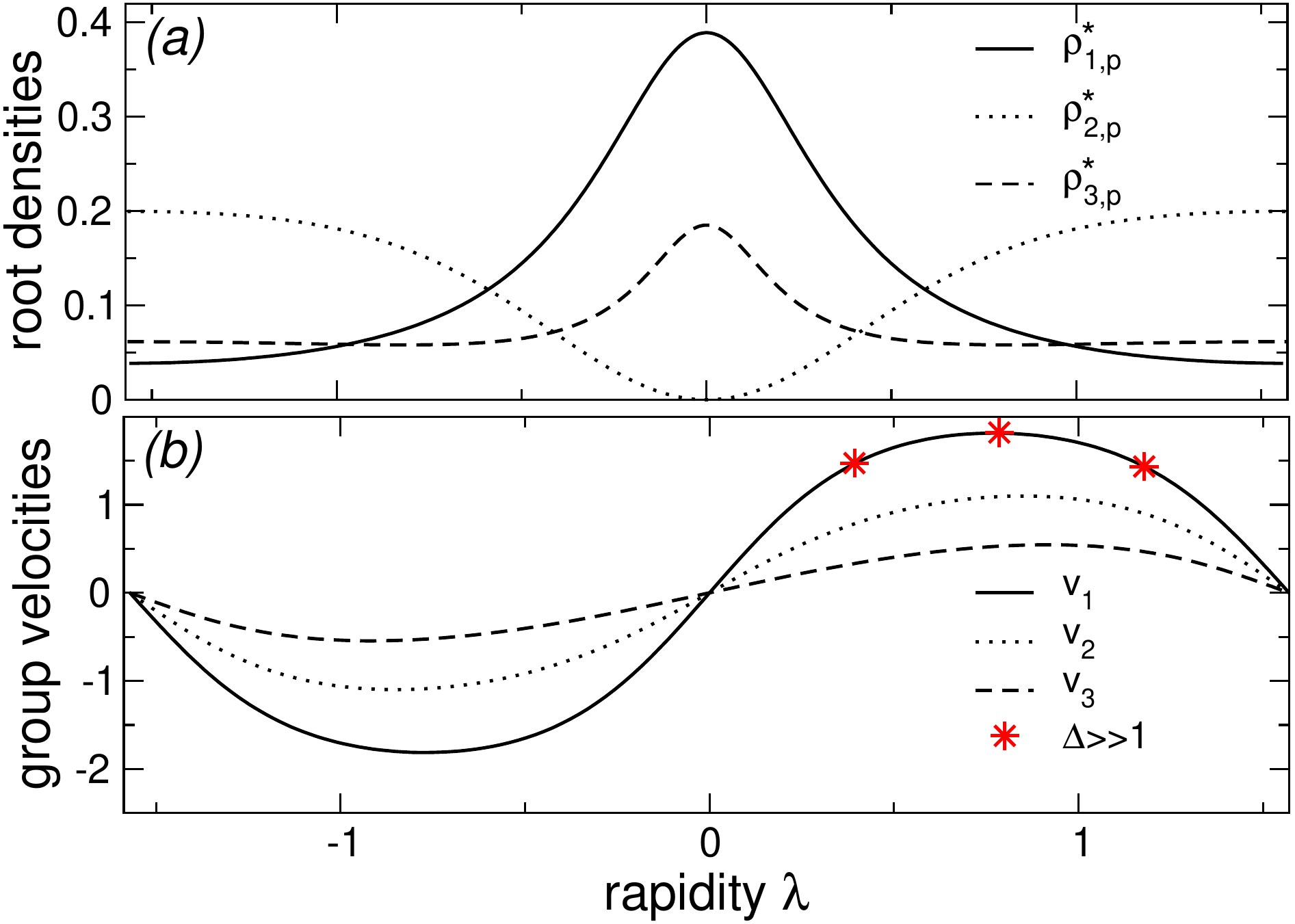}
\caption{ The post-quench steady state and the low-lying excitations around it. 
 Results for the quench from the N\'eel state in the $XXZ$ chain with $\Delta=2$. 
 (a) The macro-state densities $\rho^*_{n,p}$ characterizing the steady state. The first 
 three particle densities $\rho^*_n(\lambda)$ plotted against the rapidity $\lambda$. 
 (b) Group velocities of the low-lying excitations around the steady-state macro 
 state, as a function of $\lambda$. Note that $v_n(-\lambda)=-v_n(\lambda)$. The 
 star symbols are the results for $\Delta\gg 1$. 
}
\label{fig1-supp}
\end{figure}
%

Here we detail the calculation of the group velocities of the entangling quasiparticles, 
which explain the linear entanglement growth after the quench. 
The low-lying excitations around the post-quench steady state can be constructed as 
particle-hole excitations over the corresponding  macro-state. 
Notice that it has been verified in Ref.~\cite{de-nardis-2015} that the low-lying excitations 
around the stationary state can be used to reconstruct ``back in time'' the post-quench 
out-of-equilibrium dynamics of local observables. First, one can 
imagine choosing among the eigenstates of~\eqref{ham-xxz} a representative of the macro-state, 
identified by some BGT quantum numbers $I_{n,\alpha}^*$. Then, a particle-hole excitation, 
in each $n$-string sector, is obtained as $I^*_{n,h}\to I_{n,p}$, where $I_{n,p}(I^*_{n,h})$ 
is the BGT number of the added particle (hole). Since the $XXZ$ chain is interacting, this 
local change in quantum numbers affects {\it all} the new rapidities. The excess energy 
of the particle-hole excitation is 
\begin{equation}
\label{ph-en}
\delta E_n=e_n(\lambda^*_{n,p})-e_n(\lambda^*_{n,h}). 
\end{equation}
Remarkably, apart from the dressing of the ``single-particle'' energy $e(\lambda)$ 
(see below for its calculation),~\eqref{ph-en}  is the same as for free models. 
The change in the total momentum is obtained from~\eqref{eps} as 
\begin{equation}
\delta P=z_n(\lambda^*_{n,p})-z_n(\lambda^*_{n,h}). 
\end{equation}
Finally, the group velocity of the particle-hole excitations is by definition  
\begin{equation}
\label{group-v}
v_n(\lambda)\equiv\frac{\partial e_n}{\partial z_n}=\frac{e'_n(\lambda)}{2\pi
\rho^*_{n,t}(1+\eta^*_n(\lambda))}. 
\end{equation}
%
\begin{figure}[t]
\includegraphics*[width=0.93\linewidth]{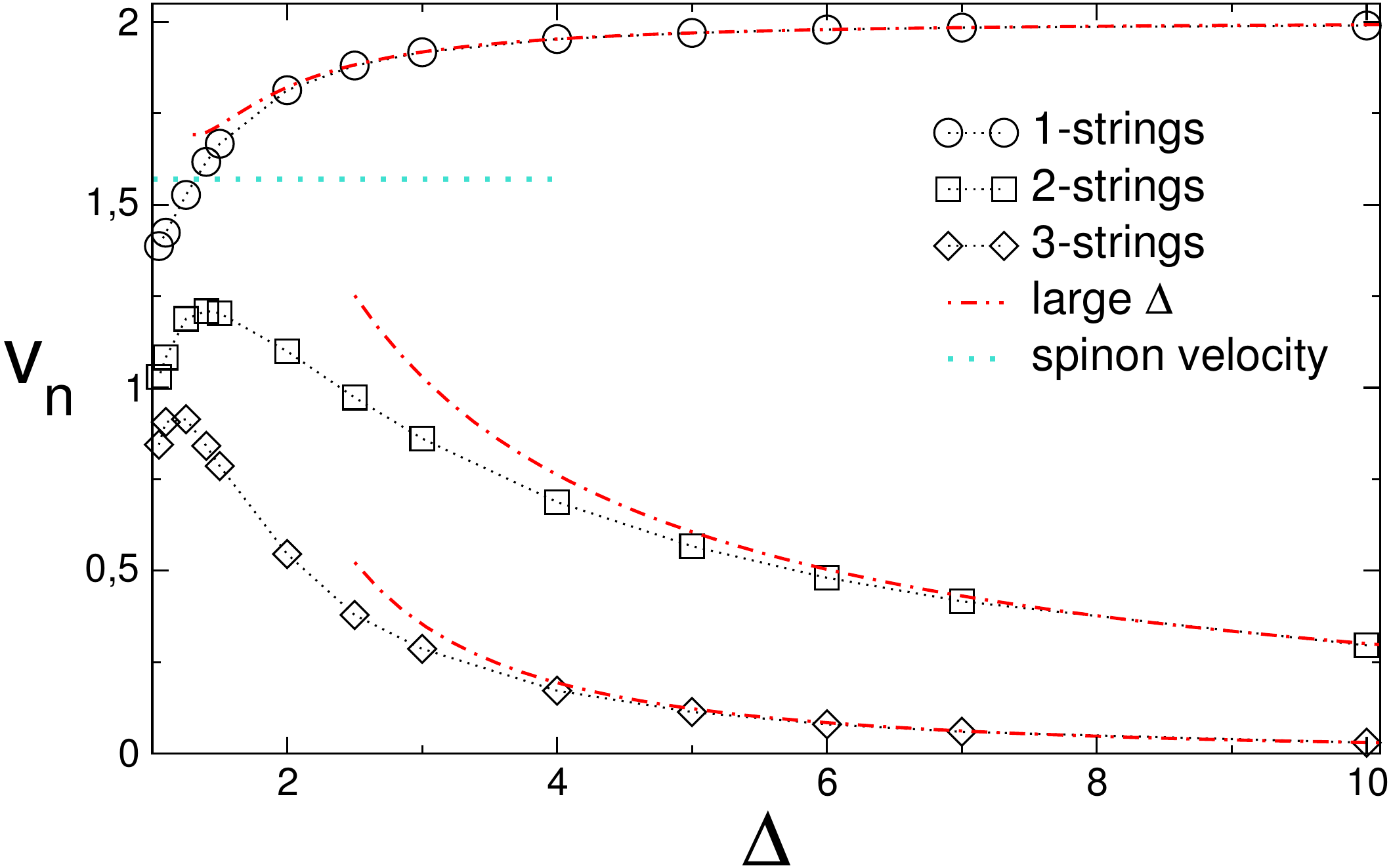}
\caption{ Group velocities of the low-lying excitations around the steady state 
 after the quench from the N\'eel state in the $XXZ$ chain. The maximum velocity in 
 each string sector (different symbols in the Fig.) is plotted against the anisotropy 
 $\Delta$. The dash-dotted lines are the analytical results~\eqref{v-pert} in the large 
 $\Delta$ limit. The horizontal dotted line is the (low-energy) spin wave velocity 
 $v_{sw}=\pi/2$ at $\Delta=1$. 
}
\label{fig-supp}
\end{figure}
%
%

Here we used that~\cite{taka-book} $d z_n(\lambda)/d\lambda=2\pi\rho^*_{n,t}$, 
with $\rho^*_{n,t}\equiv\rho^*_{n,p}(1+\eta_n^*)$, and we defined $e'_n(\lambda)$ 
as the derivative of $e(\lambda)$. Importantly, $e'_n(\lambda)$ is determined by 
an infinite system of Fredholm integral equations of the second kind as 
\begin{equation}
\label{tosolve}
e'_n(\lambda)+\frac{1}{2\pi}\sum\limits_{m=1}^\infty\int\!d\mu e'_m
(\mu)\frac{\Theta'_{m,n}(\mu-\lambda)}{1+\eta^*_m(\mu)}=
\epsilon'_n(\lambda),
\end{equation}
where $\Theta'_{n,m}(\lambda)\equiv d\Theta_{n,m}(\lambda)/d\lambda$ and 
$\epsilon'_n(\lambda)\equiv d\epsilon_n(\lambda)/d\lambda$  (cf. also~\eqref{Theta}
 and~\eqref{eps}).  The solutions of~\eqref{tosolve} can be obtained 
numerically very effectively after truncating the system considering $n\le n_{max}$. 
We should mention that the method outlined above has been used  to 
study transport properties in the $XXZ$ chain starting from inhomogeneous initial 
conditions~\cite{cdy-16,bertini-2016}, and the spreading of correlations 
after quantum quenches~\cite{bonnes-2014} (see also~\cite{calabrese-2006,calabrese-2007,gambassi-2011,delfino-2014,collura-2015,cc-16,kormos-2016,bertini-2016a,barbiero-2016,
chiocchetta-2016}). 

Importantly, in the limit $\Delta\to\infty$ the solution of~\eqref{tosolve}, and 
the group velocities~\eqref{group-v} can be obtained {\it analytically} as a power 
series in $z\equiv e^{-\eta}\approx 1/(2\Delta)$. For instance, for the quench from 
the N\'eel state one obtains 
\begin{align}
\label{v-pert}
v_1(\lambda) = & 2\sin(2\lambda)+z^2[\sin(2\lambda)+4\cos(4\lambda)\sin(2\lambda)]\\
\nonumber+&z^3[2\cos(2\lambda)\sin(2\lambda)]-z^4[2\cos(4\lambda)\sin(2\lambda)\\
\nonumber -&\sin(2\lambda)-4\cos(8\lambda)\sin(2\lambda)]+{\mathcal O}(z^5)\\
\nonumber
v_2(\lambda) = & 6z\sin(2\lambda)+{\mathcal O}(z^3)\\\nonumber
v_3(\lambda) = & 12z^2\sin(2\lambda)+{\mathcal O}(z^3). 
\end{align}
A similar result can be obtained for the Majumdar-Ghosh state as 
\begin{align}
v_1(\lambda)= & 2\sin(2\lambda)+4z^2\cos(4\lambda)\sin(2\lambda)\\\nonumber
+ & z^3[8\cos(2\lambda)\sin(2\lambda)+ 4\sin(2\lambda)]\\\nonumber
+ & z^4[6\sin(2\lambda)-4\cos(2\lambda)\sin(2\lambda)+\\\nonumber
- & 8\cos(4\lambda)\sin(2\lambda)+4\cos(8\lambda)\sin(2\lambda)]+{\mathcal O}(z^5)\\\nonumber
v_2(\lambda)= & 4z\sin(2\lambda)\\\nonumber
+ & z^2[4\sin(2\lambda)+4\cos(2\lambda)\sin(2\lambda)]\\\nonumber
+ & z^3[4\cos(4\lambda)\sin(2\lambda)-6\sin(2\lambda)]\\\nonumber
+ & z^4[4\cos(6\lambda)\sin(2\lambda)-28\sin(2\lambda)\\\nonumber
- & 32\cos(2\lambda)\sin(2\lambda)]+{\mathcal O}(z^5)\\\nonumber
v_3(\lambda)= & 12z^2\sin(2\lambda)+{\mathcal O}(z^3). 
\end{align}
As an example of calculation of group velocities we plot in Figure~\ref{fig1-supp} 
(b) $v^*_n(\lambda)$ for $n=1,2,3$ for the quench from the N\'eel state in the 
$XXZ$ chain with $\Delta=2$. In the figure, $v_n$ are plotted versus the 
rapidity $\lambda$. The star symbols are the perturbative results~\eqref{v-pert} 
in the large $\Delta$ limit. Moreover, in Figure~\ref{fig-supp} we plot the 
maximum group velocity in each string sector as a function of $\Delta$. 
The dash-dotted lines are the analytical results~\eqref{v-pert} in the limit $\Delta\to\infty$. 
Note that in the limit $\Delta\to\infty$ one has $v_1\to 2$, whereas the group velocities 
are vanishing in the other string sectors. At $\Delta=1$, $v_1$ is different 
from the low-energy spin-wave velocity  $v_{sw}=\pi/2$ (shown as horizontal dotted line in  Fig. \ref{fig-supp}).

\section{Two disjoint intervals}
\label{mi-section}

%
\begin{figure}[t]
\includegraphics*[width=0.95\linewidth]{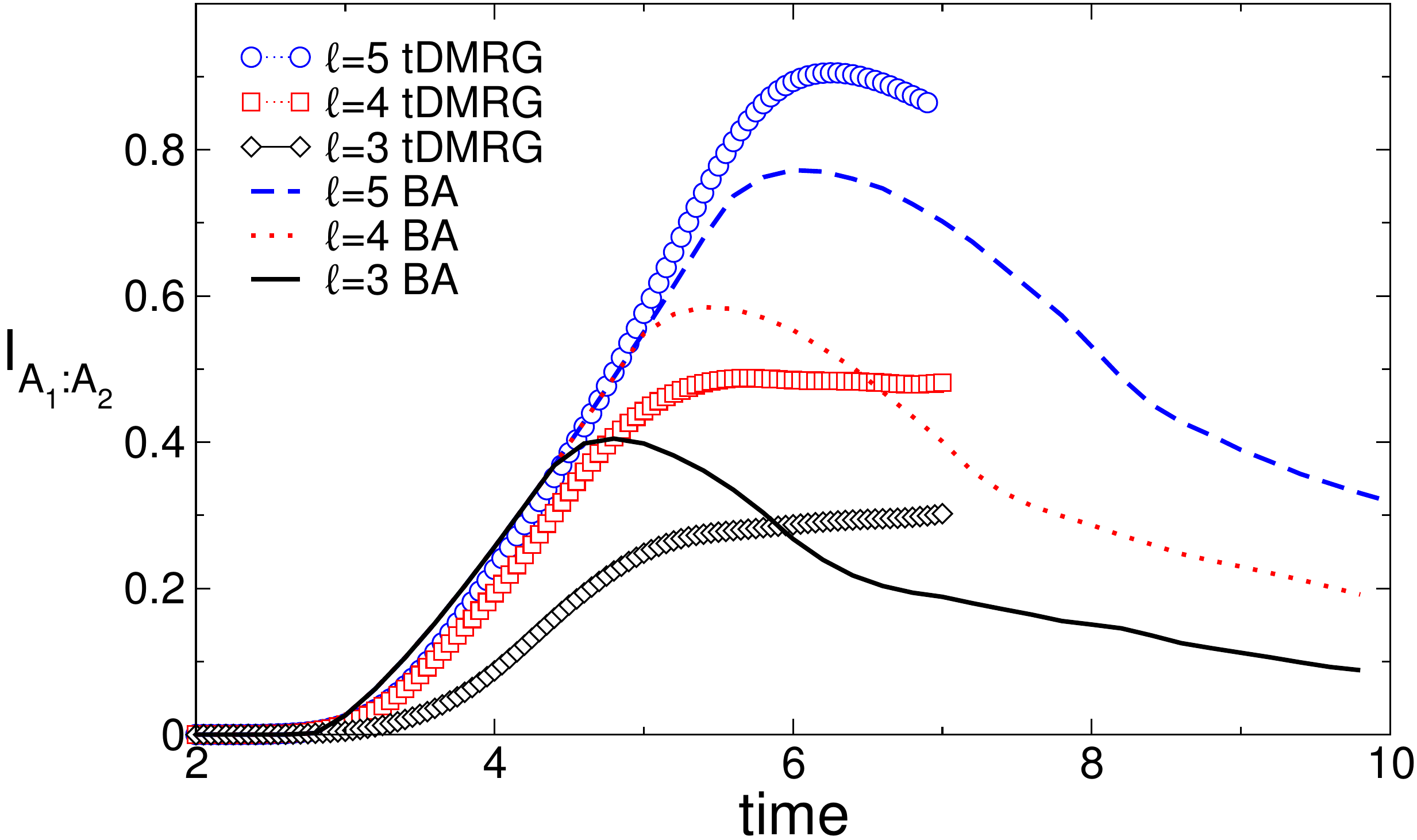}
\caption{ Mutual information $I_{A_1:A_2}$ between the two intervals 
 $A_1$ and $A_2$ after the quench from the N\'eel state in the open $XXZ$ 
 chain. Here $A_1$ and $A_2$ are at the two edges of the chain and are of 
 equal length $\ell$. The distance between $A_1$ and $A_2$ is $d=10$ lattice 
 sites. The symbols are tDMRG results for $\ell=3,4,5$. All the data are 
 for $\Delta=2$. The lines are the analytical results using the quasiparticle 
 picture \eqref{mi-quasi}. 
}
\label{fig-mi}
\end{figure}
%

In this section we investigate the entanglement of two disjoint spin blocks after a quench. 
This is motivated by some recent holographic results, which also apply to irrational $1$+$1$ CFT (see e.g. 
Refs.\cite{bala-2011,allais-2012,ab-13,lm-15,asplund-2015}), that are in contrast with the quasiparticle picture.
We focus on the behavior of the von Neumann mutual information $I_{A_1:A_2}$ after a global 
quench. We consider the tripartition of the chain as $A_1\cup A_2\cup B$, where $A_1$ and $A_2$ 
are two disjoint intervals of equal length $\ell$ and at distance $d$, while $B$ is the 
remainder of the chain. 
The mutual information $I_{A_1:A_2}$ is defined as 
\begin{equation}
I_{A_1:A_2}\equiv S_{A_1}+S_{A_{2}}-S_{A_1\cup A_2}, 
\end{equation}
with $S_{A_{1(2)}}$ and $S_{A_1\cup A_2}$ being the entanglement 
entropies of $A_{1(2)}$ and $A_{1}\cup A_2$, respectively. 

For an infinite system and two intervals of length $\ell$ at a distance $d$, it is straightforward to derive the 
contribution to the mutual information of each quasiparticle with velocity $v$, namely \cite{calabrese-2005}
\begin{multline}
\label{mi-cft}
I_{A_1:A_2}\propto -2\max((d+\ell)/2,vt)\\
+\max(d/2,vt)+\max((d+2\ell)/2,vt), 
\end{multline}
which predicts $I_{A_1:A_2}=0$ for $vt\le d/2$, a linear increase for $d/2<vt\le (d+\ell)/2$, followed by a linear decrease up to 
$vt=(d+2\ell)/2$.
In stark contrast, it has been suggested that the quasiparticle 
picture for the entanglement propagation does not hold in holographic contexts. 
The scenario of Ref.~\cite{asplund-2015} predicts $I_{A_1:A_2}=0$ at any time: 
the idea is that quasiparticles originated 
at the same point in space and travelling one in $A_1$ and the other in $A_2$ 
do not remain maximally entangled when they are far apart from each other (a phenomenon known as scrambling).

To clarify whether the quasiparticle picture applies to interacting integrable models, here we discuss the behavior of $I_{A_1:A_2}$ after the 
quench from the N\'eel state $|\uparrow\downarrow\uparrow\cdots\rangle$ in the $XXZ$ chain. 
We restrict ourselves to the open $XXZ$ spin chain. 
We always consider the situation with the two intervals 
$A_1$ and $A_2$ at the opposite edges of the chain, as it is convenient for 
the numerical simulations. This implies $L=2\ell+d$. 
In this case, the contribution of each quasiparticle is given by~\eqref{mi-cft} 
with the replacement $\ell\to2\ell$, but the formula is valid only before the revival time $t_{\rm rev}=L/v$.
Similar to~\eqref{conj}, the final quasiparticle prediction for the mutual information is obtained by summing 
(and integrating) the contribution of all quasiparticles, i.e. 
\begin{multline}
\label{mi-quasi}
I_{A_1:A_2}=\sum_n\int d\lambda s_n(\lambda) \Big[
-2\max((d+2\ell)/2,v_n(\lambda) t)\\
+\max(d/2,v_n(\lambda) t)+\max((d+4\ell)/2,v_n(\lambda) t)\Big], 
\end{multline}
which, again, it is valid before the revival time, i.e. for $t$ such that $v_M t<L$.
%
In the following we compare~\eqref{mi-quasi} with 
tDMRG results. 

It is well known that extracting the entanglement entropy of multiple disjoint 
intervals in DMRG simulations is a formidable task, in contrast with 
the single-interval entropy. Specifically, given the MPS representation 
of the state, the computational cost scales as $\chi^6$ for the multi-interval, 
whereas it is only $\chi^3$ for the single interval, with $\chi$ the 
bond dimension of the MPS. This issue is even more dramatic out of 
equilibrium, where $\chi$ grows exponentially with time. For this reason 
we can provide reliable data for $I_{A_1:A_2}$ only for very small intervals, 
and up to short times after the quench. 

Our results are presented in Figure~\ref{fig-mi}. The symbols are the tDMRG data 
for $I_{A_1:A_2}$ plotted as a function of the time after the quench. We  show the data 
for $\ell=3,4,5$, which are the only sizes we can simulate reliably, and only for 
$t\lesssim 7$. In our simulations, for all values of $\ell$, the two intervals are 
at fixed distance $d=10$. The lines in the Figure are the analytic results obtained 
using~\eqref{mi-quasi}. 
Clearly, the system sizes and time scales accessible in our tDMRG simulations do not 
allow us to reach a quantitative conclusion on the validity of~\eqref{mi-quasi} for the mutual information. 
Anyhow, we observe that the DMRG data are in a good overall qualitative agreement with the quasiparticle predictions~\eqref{mi-quasi}.
Indeed, the mutual information is zero for $t<d/(2 v_M)$, then it starts growing (seemingly) linearly with time, 
up to a maximum value which is clearly visible for $\ell=5$ (but not for smaller $\ell$).
The few results available in the literature for the mutual information of two disjoint intervals for free systems \cite{fc-10,coser-2014} 
show very similar effects when the lengths of the subsystems are very small as in our simulations. 
Hence, these DMRG results provide a strong support for the validity of the quasiparticle picture for 
$I_{A_1:A_2}$, signaling the absence of scrambling which takes place for CFTs with large central charge ~\cite{asplund-2015}. 

\bibliography{pnas-sample}

\end{document}